\documentclass[conference]{IEEEtran}
\IEEEoverridecommandlockouts
\usepackage{cite}
\usepackage{amsmath,amssymb,amsfonts}
\usepackage{algorithmic}
\usepackage{graphicx}
\usepackage{textcomp}
\usepackage{xcolor}
\def\BibTeX{{\rm B\kern-.05em{\sc i\kern-.025em b}\kern-.08em
    T\kern-.1667em\lower.7ex\hbox{E}\kern-.125emX}}
\begin{document}

\title{Human Age Estimation from Gene Expression Data using Artificial Neural Networks\\

}
\author{\IEEEauthorblockN{ Salman Mohamadi}
\IEEEauthorblockA{\textit{Computer Science and Electrical Engineering} \\
\textit{West Virginia University}\\
Morgantown, WV, USA \\
Sm0224@mix.wvu.edu}
\and
\IEEEauthorblockN{Gianfranco.Doretto}
\IEEEauthorblockA{\textit{Computer Science and Electrical Engineering} \\
\textit{West Virginia University}\\
Morgantown, WV, USA \\
    gianfranco.doretto@mail.wvu.edu}
\and
\IEEEauthorblockN{Nasser M. Nasrabadi}
\IEEEauthorblockA{\textit{Computer Science and Electrical Engineering} \\
\textit{West Virginia University}\\
Morgantown, WV, USA \\
    nasser.nasrabadi@mail.wvu.edu}
\and
\IEEEauthorblockN{Donald A. Adjeroh}
\IEEEauthorblockA{\textit{Computer Science and Electrical Engineering} \\
\textit{West Virginia University}\\
Morgantown, WV, USA \\
    donald.adjeroh@mail.wvu.edu}}

\maketitle

\begin{abstract}

    The study of signatures of aging in terms of genomic biomarkers can be uniquely helpful in understanding the mechanisms of aging and developing models to accurately predict the age. Prior studies have employed gene expression  and DNA methylation data aiming at accurate prediction of age. In this line, we propose a new framework for human age estimation using information from human dermal fibroblast gene expression data.
   First, we propose a new spatial representation as well as a data augmentation approach for gene expression data. Next in order to predict the age, we design an architecture of neural network and apply it to this new representation of the original and augmented data, as an ensemble classification approach. Our experimental results suggest the superiority of the proposed framework over state-of-the-art age estimation methods using DNA methylation and gene expression data.
    
\end{abstract}

\section{Introduction}
Healthy aging requires accurate assessment and modification of healthcare strategies and culture. However, this could not be done without understanding the aging process in terms of transcriptional changes in cellular scale. In fact, even though transcriptional changes of age-related genes differ among different tissue types \cite{b1}, modeling these changes of expression level across the human lifespan, would extend the general understanding of aging and heterogeneity in senescence of individuals with the same age. Along with this, prior studies suggest that age prediction using trasncriptional changes such as gene expression level will be feasible almost accurately and could be very helpful in disease prevention and anti-aging therapeutics.

With an emphasis on the mentioned suggestion, this work relates to the general problem of estimating human age from genomic data, a known genotype to phenotype problem \cite{b2,b3}. 
The human genome has close to 30,000 genes. Each gene responds to biological stress or defined experimental conditions, and this response is captured by its level of expression under such condition. The gene expression data records information about the observed levels of  expression for given set of genes in the genome, under specified experimental conditions. Thus, our challenge is, given a data set of gene expression values under specified experimental condition(s), in particular, gene expression data from human dermal fibroblasts, the main cell type available in skin connective tissue \cite{b4}, from several individuals, how can we recover the age for a given individual in the set?  Addressing this question will have significant implications in various fields, from health and precision medicine (chronological age and biological age) to improved understanding of experimental transcriptomic datasets. On one hand, this problem is made very challenging because of the very limited data sets currently available. On the other hand, the decreasing cost of data acquisition (and hence increasing data availability) imply that this type of problem will become a dominant feature in the health and in particular, personalized medicine of the future. This work is also expected to have tremendous impact on forensic investigations involving genomic data evidence, including human trafficking and child exploitation.


Here, first we review the prior similar approaches to human age estimation, including those that use information about gene expression, or  other types genomic data, and then we propose our framework based on a new approach to gene expression data representation and inference. 
 Our main contributions include:\\
\begin{itemize}
  \item A new framework for age estimation using a dataset of gene expression data of human dermal fibroblasts, which outperforms the state-of-the-art method on this dataset and dataset of methylation data. 
  \item  A novel data representation as well as data augmentation methods for gene expression data, which allow us to apply well-known deep learning tools such as artificial neural networks (ANNs).
\end{itemize}
 
 The paper is organized as follows. The next section reviews related work. Section 3 presents our proposed data representation scheme and data augmentation method for gene expression data. Section 4 presents our neural network-based framework for age prediction using gene expression data. Section 5 presents the experiments and the results. Section 6 concludes the paper.

\section{Background and Related Work}
With increasing availability of different types of genomic data, accurate age estimation has emerged as an important problem domain in different applications. The connection between human age and genomic data has been the focus of various studies, and has thus been investigated from many angles and perspectives. Tsuji et al \cite{b5} found some association between age and chromosome telomere lengths and used it for human age estimation. Richter et al \cite{b6}, tried to find a connection between age and mitochondria DNA while working on probable cause of high mutation rate of mtDNA. Authors of reference \cite{b7} extended the work on single tissue type to multi-tissue, hoping to represent a better age prediction model for so many tissue types. Similarly, a modeling on age based on epigenetic data is presented in \cite{b8}. With an emphasis on general pattern analysis, Peters et al \cite{b9} characterized the age by modeling variations within gene expression data inherently linked to the aging. Authors of reference \cite{b10} modified their previous model on discriminating relevant gene expression patterns related to young people and old people. Moreover, a study on whole genome methylation data by reference \cite{b11} strongly suggested general patterns of signature of aging.   Xu et al \cite{b12} proposed a method based on support vector regression to estimate the age from DNA methylation data which achieved an average accuracy of 4.7 years.
However, there are only a handful of studies that try to connect human gene expression data with age. Clearly, such a connection could have significant implications for health and disease studies, for instance, in the area of human biological age estimation \cite{b13,b14}, or in personalized medicine. In other words, the problem of age estimation can be studied using the increasingly available gene expression data, with a potential for significantly improved accuracy at  low cost.

Holly et al \cite{b10} provided a statistical model based on gene expression data for age-group classification which helped pave the way for this avenue of research.
Fleischer et al \cite{b15} developed a computational method to predict age based on gene expression data from skin fibroblast cells using an ensemble of machine learning classifiers. They generated an extensive RNA-seq dataset of fibroblast cell lines derived from a total of 143 individuals whose ages range from 1 to 94 years. Using this dataset, their method predicted chronological age with a median error of 4 years, outperforming algorithms proposed by prior studies that predicted age from DNA methylation \cite{b12,b17,b10}, and from gene expression data  for fibroblasts  \cite{b10}. 


This work is along this same line of estimating age for individuals using information from their gene expression data. Our work is different in terms of the new data representation scheme we introduce, and the proposed new data augmentation method for this type of data. With these two innovations, it becomes possible to build a parallel framework that can exploit the power of deep learning techniques, such as artificial neural networks. While there are also machine learning and deep learning tools for 1-D data representation that could be used for these types of data, our 2-D data representation achieves the highest accuracy compared to previous work. More importantly, one of the key advantages of the  spatial (2-D) representation is that, due to the fixed spatial location of expression value associated with each gene, there is more flexibility in exploiting potential long-range associations between genes \cite{b18,b19,b20}, beyond just adjacent or co-located genes. 

\begin{figure}[t]
\begin{center}
\includegraphics[width=1\linewidth]{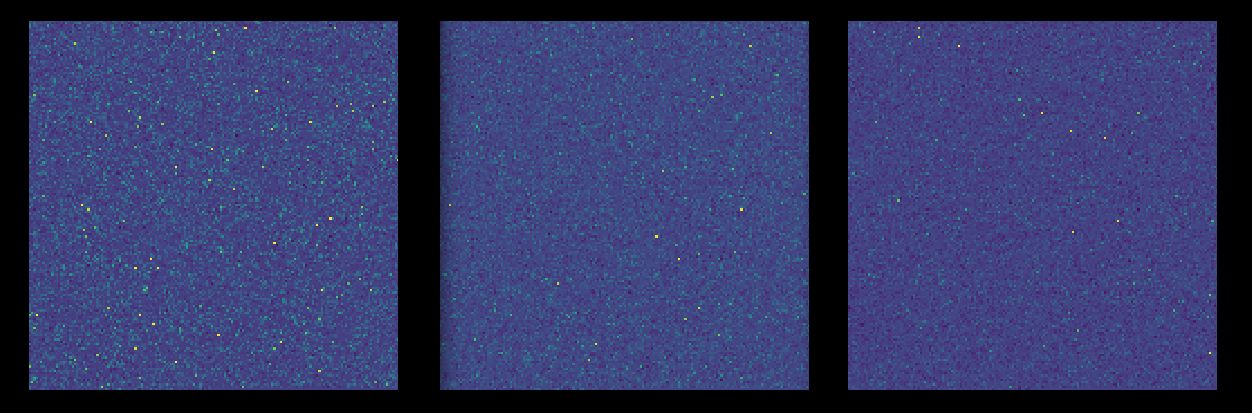} \rule{0.1\linewidth}{0pt}
   
\end{center}
   \caption{Spacial data representation for three individuals; left to right: 1 year old, 30 years old and 61 years old.}
\label{fig:001_SDR}
\end{figure}

   

\begin{figure}[t]
\begin{center}
\includegraphics[width=1\linewidth]{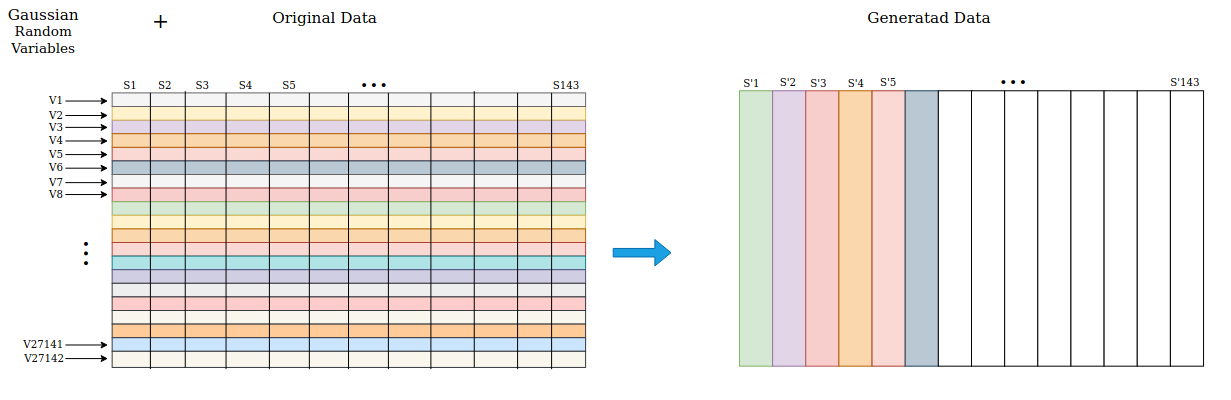} \rule{0.1\linewidth}{0pt}
   
\end{center}
   \caption{Data Augmentation Scheme.}
\label{fig:001_DA}
\end{figure}

\section{Data Representation and Augmentation}
In this section we describe our novel data representation scheme and data augmentation approach for gene expression data which enable us to use artificial neural networks.

\subsection{Dataset and Normalization}
Our  database which is from \cite{b15}, consists of a matrix of expression values from 27,142 genes for a total of 143 individuals represented in two groups (133+10). Age ranged from 1 year to 94 years. In fact, the dataset provides us with expression value of each and every gene for all individuals. Certainly, with only 143 samples, there is need for more data in order to train our neural network. Thus, we need to perform data augmentation to generate synthetic datasets.
Each row contains the expression values from certain genes across 143 individuals. Each column is the expression values of the 27,142 different genes for a given subject. Considering the rows, it seems that despite the values in a column, values in a row are in a smaller range, i.e., expression values for a certain gene across all individuals are  in a relatively small dynamic range, but different genes expression values for  a certain individuals vary significantly. This has also been observed in other general genomic datasets. \cite{b21,b22,b23}.

Data normalization in known to accelerate the learning process of the network. Thus, before doing any further steps such as data augmentation, we normalize each row (specific gene) of the database individually, as a pre-processing step. After applying standard normalization to each row, as shown below, corresponding data samples of each row are centered at zero with  unit variance: ${z = \frac{(x - \mu)}{\sigma},}$
where ${\mu}$ and ${\sigma}$ are respectively, the mean and standard deviation of the original data samples $x$, and $z$ is the normalized data.
\subsection{Spatial Representation}
Here, we propose a new data representation for the gene expression data from  an individual, and a methodology for data augmentation. In fact,data representation and in general data visualization is very helpful in better extracting the necessary information or modeling the data \cite{b38,b33}. To analyse the data, we need to extract the columns of the data matrix, which captures the expression values across all genes (27,142 genes), for each individual. 
As we will explain in the next section, we design a deep neural network classifier in order to perform the classification w.r.t the age for the subjects. We extract each column and reshape it to a square two-dimensional matrix, where each element of this matrix represents the expression values for a specific gene. We reshape the vector by moving the elements in the vector to a the matrix, filling from top left to the down right of the matrix (165x165=27,225) which the last 83 elements of the matrix (27,225-27,142=83) are padded with zeros.
Therefore, we will have 143 squared matrices for the 143 subjects. Then we transform each matrix to a square image data, where each pixel location of every image represents the gene expression value of a specific gene. By so doing, we provide a spacial (2-D) representation for expression values of all genes for each individual. Fig.  \ref{fig:001_SDR} shows some examples for three individuals in different age groups. 
More importantly, the figure also shows the nature of the 2D representation for the gene expression data, and the difficulty of analyzing such representations. It is noteworthy to mention that while one could say that representing 1-D data in the form of 2-D data may seem ad hoc (leading to possibly irrelevant spatial relationships), this type of representation makes it simpler to explore and possibly exploit potential long-range relationships between genes, for instance, using simple 2D convolution kernels. This may not be as simple, or as computationally efficient, using 1-D representations.   
Further, since each pixel location in this type of 2-D data (image) is associated with a unique gene, this could be even helpful at the analysis stage (for instance, using artificial neural networks) in terms of allowing us to easily compare a given gene across all individuals (or subset of individuals) in the data set.


\subsection{Data Augmentation}
\label{sec:data_augment}
Although the data representation is a 2D matrix of each column's data samples, the resulting images are not the same as traditional natural images, like face, or natural real-world objects. First, based on our pre-evaluation experiment, well-known data augmentation techniques for image data, such as basic image manipulations \cite{b24}, did not prove to be effective for this type of image data. The reason may be the nature of the data, which is totally different from normal image datasets. Further, in applying data augmentation on this type of data, we need to recognize the meaning of the resulting augmented data. For instance, simply flipping the image, or rotating the data will lead to outcomes that may not be meaningful, given the information represented in the images. 

Thus, in order to augment this type of data, we should consider the basic representation of the data, in other words, augment the data in the feature space. In this line, besides machine learning tools, classical signal processing methods and tools could be effective \cite{b28,b30,b37}. We need to generate new features (expression values of a given gene, which is in a row) for a certain subject (values in a given column), without significantly shifting the statistical distribution of the sample data\cite{b25} to generate a new subject with the same label (no change in the subject's label),

Therefore, as shown in Fig. \ref{fig:001_DA} and presented in corresponding equations, first we add a random Gaussian variable with mean zero and standard deviation ${\sigma}$=${\beta}$*${\sigma_d}$, where ${\sigma_d}$ is the standard deviation of the original data samples of a given row; then allocate the same label of the original data (age of each column as an individual) to the columns of the generated data. Here ${\beta}$ is an adjustable parameter (based on our experiments, the best value for ${\beta}$ is 0.13). Here we have ${X \sim \mathcal{N}(\mu,\,\sigma^{2})\,}$ and ${P(x) = \frac{1}{{\sigma \sqrt {2\pi } }}e^{{{ - \left( {x - \mu } \right)^2 } \mathord{\left/ {\vphantom {{ - \left( {x - \mu } \right)^2 } {2\sigma ^2 }}} \right. \kern-\nulldelimiterspace} {2\sigma ^2 }}}}$.


In other words, we need to add some random variable to each row (feature) considering the standard deviation of the row, without significantly changing the data distribution. Otherwise we can not assign the label of the original data to the generated data. Ideally, in order to be able to assign a label to a generated data which is not the same as the label of the original data, we need to comprehensively model the biological functionality and role of each gene in aging. However, this is not practical with today's technology. Therefore, our approach to augment the data is to generate new data samples which do not remarkably deviate from the original data samples distribution individually (just a new copy of the data samples with a small deviation) so that we can assign the same label to the generated data sample. While this may be seen as a strong assumption from one viewpoint, this has already been used by other data augmentation methods for different data types \cite{b26,b27}. Our results show that this assumption is sound.
Using this method we will synthesize different versions of the dataset with the corresponding labels  \cite{b25,b27,b36}. 

We have tried several values for  ${\beta}$, starting from 0.05 to 0.25. Using the validation data, the best results is associated with ${\beta=0.13}$. 
Since we have 27,142 different genes, we need to use 27,142 random
random Gaussian variables with mean zero and standard deviation associated with the standard deviation of each row and add them to the data accordingly, in order to generate a new set of the data with the same labels.
For data augmentation, since we added a random variable to each row for every set for data augmentation, one may ask why just 5 sets of augmentations? In order to address this question, we have also performed 5 more sets (totally 10 sets) of data augmentation. However, the results did not improve remarkably. We suspect that this may be caused by the fact that we may be generating similar copies of the augmented data when we generate many sets.

As discussed later in Section 5, we need to design our experiments in such a way that divides the dataset into several non-overlapping parts, where each part should be independently used for training and testing in different rounds of the experiment. We use a major proportion of the original data for training and the remaining part as the test data. Data augmentation is performed only on the training data. For data partitioning, the data is randomly divided into four fixed partitions, then at every round of the experiment (four rounds in total), we use three out of four parts for the data augmentation. Here, all these three parts (${75\%}$ of the original data) plus augmented data are designated for training, and the last part $(25\%)$ is used for testing. Furthermore, we mention that, data augmentation as described above is applied five times to the ${75\%}$ of the original data.  Thus, at each round, the training data will consist of 107 + (5 x 107) = 642 images, while the test data consists of 36 images.

\section{Age Estimation Framework}
In this section we present our framework for age estimation from gene expression data that exploits the outcome of our proposed data representation and data augmentation schemes. After data representation and augmentation, we feed the data to a network designed for age-group classification. Given our 2-D representation of the gene expression data, it become natural to analyze such images using two dimensional convolutional neural networks (CNNs). We emphasize that our data representation scheme (spatial 2-D rather than 1-D representation) makes it natural to exploit the modeling capability of  deep learning architectures for the problem of age estimation using our type of data. In fact, with our spatial representation, each pixel location represents a specific gene, while the pixel intensity value indicates the expression level for the gene. Thus, any specific gene is located in a fixed pixel location in the image, possible relations between genes can be captured using a simple convolution mask. This means that our neural network can more easily learn these possible relationships, even between genes that may be widely separated in the genome. This is confirmed in our experiments by changing the order of genes in the spatial representation for some of individuals and then training the network. In fact, one possible reason could be that even by changing the order of the genes, almost for all individuals in a certain age group, the expression values are quite similar.

\begin{figure}[t]
\begin{center}
\includegraphics[width=.8\linewidth]{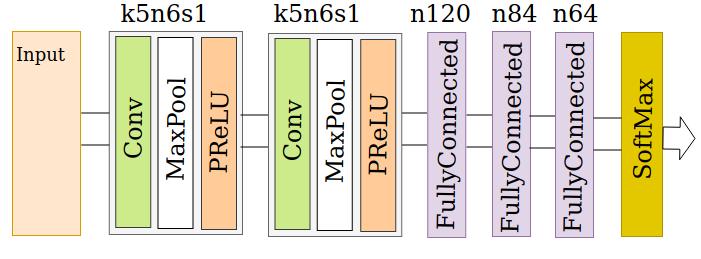} \rule{0.1\linewidth}{0pt}
   
\end{center}
   \caption{Network architecture for age estimation via an ensemble of age-group classifiers.}
\label{fig:001_CNN}
\end{figure}

\subsection{Network Architecture}
The core objective in this work is to develop an effective framework for age estimation from gene expression data. However, as mentioned earlier, in order to accomplish it, we need a suitable data representation, and a proper method for data augmentation, especially given the limited data sizes. Given our 2-D spatial data representation, artificial neural networks (ANNs), for instance, using convolutional neural networks (CNNs) become natural candidates for learning in such a framework. 
Here, we elaborate on our network architecture and the rationale behind its design. Initially, we performed pre-evaluation experiment on some well-known deep neural networks, such as VGG19 and ResNet18 in order to emphasize the importance of compatibility between the network and the data or task \cite{b39,b34}. Our preliminary experiments showed that, due to the size of dataset, these deep neural networks did not produce good results. Thus, for this work, we designed a shallower neural network as shown in Fig. \ref{fig:001_CNN}. 
The first two layers are two identical convolutional blocks consisting of a convolution layer followed by max-pooling (2 x 2) and ReLU activation function. Then we incorporate three fully connected layers and finally a softmax classifier as the last layer of the network.

\subsection{Age Grouping}
\label{sec:age_grouping}
One major problem of age estimation from genetic datasets is the paucity of data with age information. Our gene expression dataset has only 143 individuals, which is quite small for training a neural network. One way to mitigate this problem is to use the data for classification of age groups, rather than direct age estimation, which will be a regression problem. Combining the classification results, one can then perform the required age estimation. 
Previous work on this dataset \cite{b15}, proposed several ensemble classifiers using different machine learning tools.
The best median absolute error and mean absolute error reported were 4 and 7.7 years, respectively. Inspired by this work, we developed our framework for ensemble classification using neural networks on our proposed data representation and augmentation. This resulted in an average median absolute error and mean absolute error of 4.1 and 3.69 years, respectively. Fleischer et al \cite{b15} incorporated twenty sets of age groupings each with a {\it fixed} age bin widths of 20 years. Working with so many age groups will obviously be computationally intensive, since we have to perform classification for each grouping. In our work, we use just six sets of age groupings,  with possibly varying age bins, and bin sizes. Given the small size of our dataset, the first sets of age-groupings used bin widths of 20 years, but varying bin positions. This makes it easier for each age group to have enough subjects for the system to learn during training. The last two sets of groupings are defined based on biological rationale \cite{b29}. The age ranges, size and structure of the six age-groupings (G1 to G6) are shown below:

\noindent 

G1: [0-20], [21-40], [41-60], [61-80], [81-100] \\  
G2: [0-5], [6-25], [26-45], [46-65], [66-85], [86-100]\\ 
G3: [0-10], [11-30], [31-50], [51-70], [71-90], [91-100]\\ 
G4: [0-15], [16-35], [36-55], [56-75], [76-95], [96-100]\\ 
G5: [0-2], [3-12], [13-24], [25-45], [46-64], [96-100]\\ 
G6: [0-11], [12-24], [25-49], [50-69], [70-100]\\ 

\begin{figure*}[t]
\begin{center}
\includegraphics[width=1\linewidth]{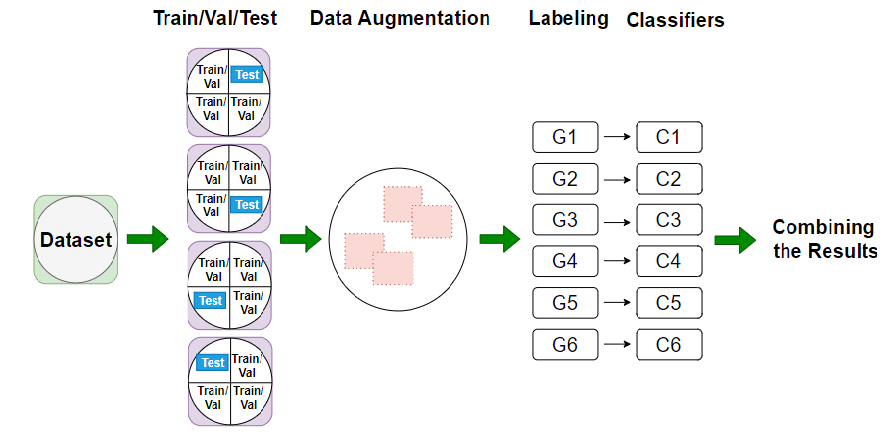} \rule{0.1\linewidth}{0pt}
   
\end{center}
   \caption{Pipeline of the proposed framework for age estimation from gene expression data.}
\label{fig:001_PiL}
\label{fig:PiL}
\end{figure*}

\subsection{Combining Classifiers Results}
Each classification round associated with one of the train/test partitions, predicts an age group for the test subjects individually. Here, the label with the highest probability in the output of the softmax is selected as the predicted class or age-group.  We need to combine the results to compute an estimated age for each subject and then average the results over the four rounds. Our approach for combining the results is similar to voting-based regression. 

Suppose that for a given test subject ${S_x}$, the six classifiers predict the following intervals as their respective age bins:  ${I_1}$, ${I_2}$, ${I_3}$, ${I_4}$, ${I_5}$ and ${I_6}$; Let these intervals be represented by their corresponding start and end points: $<s_i, e_i>$, $i=1,2,...,6$. To compute a single number for the estimated age rather than an age bin, first, we compute the mean of the intervals, next we compute the average of all six means as the estimated age for ${S_x}$. That is, 
$A_x=\frac{1}{6}{\sum_{i=1}^{6} (\frac{e_i+s_i}{2})}$, where $A_x$ is the estimated age for subject $S_x$.  

Besides the accuracy of the individual classifiers, as a metric of evaluation, we also used the mean absolute error, and the median absolute error.

\section{Experiments}

\subsection{Data Augmentation} 
As described in the section \ref{sec:data_augment}, we add a random Gaussian variable with mean zero and adaptive standard deviation to each row. We have ${X \sim \mathcal{N}(\mu,\,\sigma^{2})\,}$
where ${\sigma}$=${\beta}$*${\sigma_d}$ and ${\sigma_d}$ is the standard deviation of the original data samples of a given row (each row consists of the expression values of a certain gene across all individuals).


Next we assign the same label of the columns of the original data to the columns of the generated data (each column contains the expression values of all genes for a certain individual), where ${\beta}$ is a parameter adjusted using validation data. Since we use the label of the original subject for the generated sample subject, we need to adjust the ${\beta}$ so that the new generated subject does not deviate notably from the original sample subject. Therefore, we should try several  values of ${\beta}$ and test the accuracy of the classifiers (mean absolute error) over the validation data. 
We have tried several  ${\beta}$ values, starting from 0.05 to 0.25. We found that best results were associated with ${\beta=0.13}$. 

\subsection{Framework Implementation}
We carried out a set of parallel experiments as depicted in Fig. \ref{fig:001_PiL}. Specifically, we randomly divided the original dataset (143 columns associated with the gene expression values of 143 subjects) into four fixed parts. For each experiment, one of the parts (${25\%}$) was selected as the test data and the remaining three parts ($75\%$) were used for data augmentation and training; which is divided to two parts, one for training (${65\%}$) and one for validation (${10\%}$). The validation data is used for hyper-parameter tuning and also adjusting the value of ${\beta}$. 

As we have six sets of age-groupings, at each of the experiments we assigned corresponding labels to each subject (image) based on the labeling protocol presented in Section \ref{sec:age_grouping}. Then we trained six identical classifiers with the proposed architecture as shown in Fig. \ref{fig:001_CNN}, each of them for one of the labeling protocols. At the end of the experiments, we would have four sets of test results, with each result associated with one of the four independent rounds of the experiments (train/test partitions in Fig. \ref{fig:001_PiL}), where each set of test results provides us with six predicted labels (one label from each of the six classifiers). The results of the six classifiers are combined to estimate of the subject's age.
We used Adam algorithm \cite{b35} as the optimizer, using 100 epochs, and a learning rate of 0.001. Our loss function was the cross entropy loss, and we wrote the code from scratch in Pytorch. 


In binary classification, where the number of classes $N$ equals 2, the cross-entropy loss can be calculated as:
\begin{equation}\begin{split}
\mathcal{L}_{C}=-(y\log(p) + (1-y)\log(1-p)).
\end{split}
\label{20}
\end{equation}

However, in this work we deal with a multiclass classification scheme where the number of classes, $N$, can be 5, 6, or 7 depending on the corresponding age grouping. Thus, we have:
\begin{equation}\begin{split}
\mathcal{L}_{C}=-\sum_{c=1}^{N}(y_{o,c}\log(p_{o,c}))
\end{split}
\label{21}
\end{equation}

where, $N$ is the number of classes, and $y$ is the binary indicator function representing whether the output observation is the ground truth class. Likewise, ${y_{o,c}}$ is defined for the multiclass scenario, and ${p_{o,c}}$ represents the probability of the output being ground truth class.
\section{Results}
\subsection{Age-group classification results}
Before proceeding with age estimation, we first investigated the performance of the proposed data representation and network on age group classification, using each of the six age groupings, since these will form the basis for our age estimation. Table \ref{ageClassify} shows the classification results for each of the six sets of age-groupings. Table \ref{ageClassify} shows the classification performance on each age-grouping, using each of the four data partitions. 
The best accuracy is associated with the classifier ${\#1}$ and ${\#6}$ respectively. Since the label data is balanced, this may imply that these two age groupings are possibly more effective  for age prediction from a biological perspective. This may be because these two groupings were derived based on some biological considerations \cite{b29}.


\begin{table}
\begin{center}
\begin{tabular}{|p{1.3cm}|p{1.3cm}|p{1.3cm}|p{1.3cm}|p{1.3cm}|}
\hline
Classifier &  Accuracy ($P_1$)  &  Accuracy ($P_2$) & Accuracy ($P_3$)  & Accuracy ($P_4$) \\
\hline\hline
 C1  &  ${100\%}$ & ${100\%}$ &  ${97\%}$ &  ${100\%}$  \\ \hline
C2 &  ${69\%}$ & ${69\%}$ &  ${72\%}$ &  ${72\%}$ \\
\hline
 C3  &  ${75\%}$ & ${72\%}$ &  ${75\%}$ &  ${75\%}$\\
\hline
 C4  &  ${78\%}$ & ${81\%}$ &  ${78\%}$ &  ${81\%}$ \\
 \hline
  C5 &  ${64\%}$ & ${61\%}$ &  ${64\%}$ &  ${61\%}$\\
\hline
  C6 &  ${89\%}$ & ${89\%}$ &  ${92\%}$ &  ${92\%}$  \\
\hline
\end{tabular}
\end{center}
\caption{Accuracy on different classifiers/age-groupings (see also Fig. \ref{fig:001_PiL}). $P_1, P_2, P_3$ and $P_4$ denote the results for experiments using train/test data partitions $P_1, P_2, P_3$ and $P_4$, respectively. The best accuracy is associated with the classifier ${\#1}$ and ${\#6}$ respectively.}
\label{ageClassify}
\end{table}

\subsection{Age estimation results and comparison}
\begin{figure}[t]
\begin{center}
\includegraphics[width=.9\linewidth]{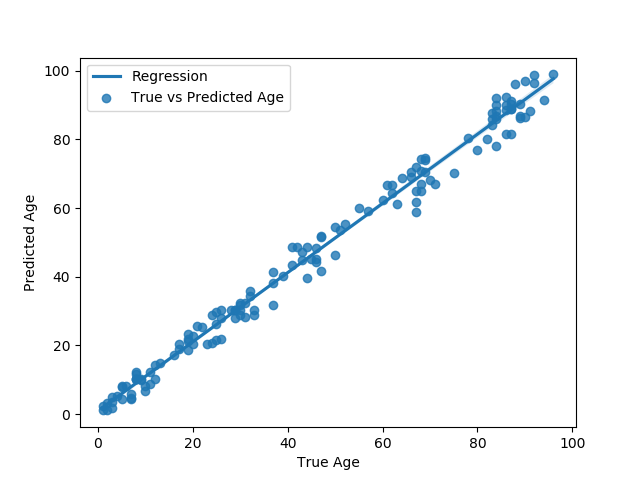} \rule{0.1\linewidth}{0pt}
   
\end{center}
   \caption{True vs predicted age.}
\label{fig:001_TvsP}
\end{figure}

To ensure robustness, we expose our framework to more randomness, by performing the whole experiment three times. 
Fig. \ref{fig:001_TvsP} shows the plot of the true age compared with the estimated age using our proposed method.
The estimated ages are generally close to the true ages.
Table \ref{ageClassify} summaries the performance of the proposed framework. The table shows the impact of the 2-D representation used in this work, and of the data  augmentation. 
Using the performance metric of mean absolute error (MAE), our proposed method outperforms current state-of-the-art method on this dataset  \cite{b15}, and is comparable to similar methods on other genomic datasets, such as epigenetic data \cite{b8}, or DNA methylation data \cite{b31}. Our method produced comparable results in terms of the median absolute error (MdAE).


\begin{table}
\begin{center}
\begin{tabular}{|p{0.6cm}|p{2.2cm}|p{1cm}|p{1.cm}|p{1.cm}|}
\hline
Ref. & Algorithm & Age bin &  MAE   &  MdAE \\
\hline\hline
 \cite{b15} & LDA ensemble  &  20  &7.7 &  4  \\ \hline
 \cite{b15} & GNB ensemble & 20 & 16 & 8 \\
\hline
 \cite{b15} & RF ensemble  &  20 & 11.8 & 5 \\
\hline
  \cite{b15} & SVM  & N/A & 11.9  & 10.2\\
\hline
\hline
Ours &   1-D ANN (no DA) & varying & 11.8  & 9\\
\hline
Ours &  ANN (2D, no DA) & varying & 7.1  & 7.7\\
\hline
Ours &  ANN (2D, DA)  & varying & 3.69  & 4.1  \\
\hline
\end{tabular}
\end{center}
\caption{Accuracy of age estimation from fibroblast transcriptomes, for various algorithms. Acronyms: MAE (Mean absolute error); MdAE (Median absolute error); GNB (Gaussian naive Bayes); SVM (Support vector machines); RF (Random forest); DA (data augmentation). }

 \label{res_ageEstimation}
 \end{table}
 
In fact, compared to the reference \cite{b15}, it can be seen that our three key contributions in this work, i.e., data representation, data augmentation, and new framework for prediction, have helped to improve the prediction accuracy. Without data augmentation it was not feasible to achieve better results using the neural network. This is because almost all of the deep learning tools, and in particular neural networks, need huge amounts of data to train\cite{b16,b32}. Moreover, our initial experiment without data augmentation did not lead to desired training of the artificial neural networks, and consequently the results were not satisfactory as it is reported in  Table \ref{res_ageEstimation}. 
 Our proposed framework achieved a mean absolute error of 3.69 and median absolute error of 4.1 which shows the effectiveness of the method compared to the results of the work on gene expression data such as \cite{b15,b9,b10} as well as work on methylation data such as \cite{b7,b11,b12}

\section{Conclusion}
Human age estimation using genomic data has important implications for research on multiple areas including understanding the individual aging, relevant healthcare strategy and precision medicine regarding the aging.

 In this paper, we proposed a new framework for age estimation using a dataset of gene expression values, for subjects with age ranging from 1 year to 94 years. The core idea of the proposed framework is to use a new representation of the information in the gene expression data, along with appropriate data augmentation for this type of data to improve the accuracy. This is then accompanied with a suitable neural network architecture to more accurately estimate the age. Experiments using the proposed framework show the effectiveness of the proposed approach. A comparative analysis with state-of-the-art methods in age estimation using genomic data is also included. 


\end{document}